\documentclass[useAMS,usenatbib,usegraphicx]{mn2e}
\usepackage{epsf}

\title[A Realistic Model for Halo Substructures]
{A Realistic Model for Spatial and Mass Distributions of Dark Halo 
Substructures: An Analytic Approach}
\author[M. Oguri and J. Lee]
{Masamune Oguri\thanks{E-mail:oguri@utap.phys.s.u-tokyo.ac.jp} 
and Jounghun Lee\thanks{E-mail:lee@utap.phys.s.u-tokyo.ac.jp} \\
Department of Physics, School of Science, University of Tokyo,
Tokyo 113-0033, Japan}
\begin{document}

\date{\today}

\voffset- .65in

\pagerange{\pageref{firstpage}--\pageref{lastpage}} \pubyear{2004}

\maketitle

\label{firstpage}

\begin{abstract}
We construct a realistic model for the dark halo substructures, and
derive analytically their spatial, mass, and velocity distributions. 
Basically, our model is a modification of the Press-Schechter theory 
to account for the dominant dynamical processes that mark the evolution
of dark halo substructures such as the tidal stripping and dynamical
friction. Our analytic model successfully reproduces all the well known 
behaviors of the substructure distributions that have been found in
recent numerical simulations: the weak dependence of the mass
distributions on the host halo mass; the anti-bias of the spatial 
distribution relative to the dark matter particle components; 
the nearly power-law shapes of the mass and velocity distributions. 
We compare our analytic results with recent high-resolution 
N-body simulation data and find that they are in excellent agreement
with each other. 
\end{abstract}


\section{INTRODUCTION}

The substructures of dark matter halos have
attracted many attentions, especially because of a possible problem they
poses in the currently popular cold dark matter (CDM) model. 
Numerical simulations have confirmed that the CDM model predicts roughly 
10\% of mass in a dark halo is bound to substructures 
\citep*{tormen98,klypin99,okamoto99,moore99,ghigna00,springel01,zentner03,
delucia04,kravtsov04}.
Although the current standard CDM paradigm has been very successful in
matching the observational data on large scale, it has been noted that
the CDM model overpredicts the abundance of substructures compared with
that of observed galactic satellites 
\citep*[e.g.,][]{kauffmann93,moore99}. This CDM problem on sub-galactic
scale is regarded as one of the most fundamental issues that has to be 
addressed. 

It has been proposed that this CDM problem on sub-galactic scale could
be resolved by changing radically the nature of dark matter. 
The proposals include the self-interacting dark matter model 
\citep{spergel00}, the warm dark matter model \citep*{colin00,bode01}, 
the annihilating dark matter model \citep*{kaplinghat00}, and
non-thermally produced dark matter \citep{lin01}. 
Another possibility is to introduce new inflationary models that can 
produce density fluctuations with small-scale power cut-off such as 
an inflation model with broken scale-invariance \citep{kamionkowski00} 
and a double hybrid inflation \citep{yokoyama00}.  
However, these rather radical models are gradually in disfavor with recent 
observations and numerical simulations \citep[e.g.,][]{hennawi02,yoshida03}.

The problem may be resolved also by taking account of astrophysical
processes such as photoionizing background \citep{somerville02} and
inefficient star formation in small mass halos \citep{stoehr02}.
These ideas claim that the observed number of satellite galaxies is small
because only very massive substructures contain stars and most
substructures are {\it dark}. To detect these small dark substructures, 
the gravitational lensing can be a very powerful tool.
For instance, we can constrain the amount of substructures through
anomalous flux ratios \citep{mao98,metcalf01,chiba02,dalal02},
spectroscopy \citep{moustakas03}, or monitoring \citep*{yonehara03}.
Although the gravitational lensing allows us to probe the distribution
directly, a caveat is the result is dependent on the spatial
distribution of substructures \citep*{chen03,evans03}. 
However, these ideas based on the dark substructures may not be
consistent with observations, either: Recent high-resolution  
numerical simulations have found that the massive substructures tend 
to place in the outer part of host halos, which is not the case for 
the satellites of our Milky way \citep*{delucia04,taylor04}.  

Despite the importance of substructures to understand the structure
formation in the universe, most work had to resort to numerical
approaches in studying mass and spatial distributions of substructures.
Compared with the numerical approach, the analytic approach has an
advantage that it allows us to compute distributions in a wide range of
parameters, e.g., mass of host halos and power spectrum. This is
essential in cosmological applications of substructures, e.g., in
computing power spectrum from halo approach \citep*{sheth03b,dolney04}.  
In contrast, in N-body simulations one needs high force and mass
resolutions to overcome ``the overmerging'' problem \citep{klypin99},
which makes the large number of reliable calculations difficult.

There are several attempts to derive the mass function of substructures
by using analytic methods \citep[e.g.,][]{fujita02,sheth03a}. 
These previous approaches, however, were all based on the oversimplified
assumptions such as the mass conservation and the uniform spatial
distribution of dark halo substructures. They considered only the
gravitational merging of substructures, without taking account the
effects of other important dynamical processes that drive the
substructure evolution, leading to their final distributions. Among them,
most important are the mass-loss caused by the global tides of host
halos \citep[e.g.][]{okamoto99}, and the orbital decay driven by 
the dynamical frictions \citep[e.g.,][]{tormen98,kravtsov04}. 

Very recently, \citet{lee04}, for the first time, developed an analytical
formalism for the global substructure mass function by incorporating the
effect of tidally driven mass-loss effect and the non-uniform spatial
distribution of substructures in dark halos. Yet Lee's approach was
still a simplification of reality, using a crude tidal-limit approximation
in estimating the tidal mass-loss and also ignoring the effect of
dynamical friction which may be more important in estimating the
abundance of massive substructures \citep{hayashi03}. 

In this paper, we construct a more realistic model for the evolution of
dark halo substructures than previous approaches by taking account of
not only the effect of global tides but also the effect of dynamical friction,
and derive analytically the spatial, mass, and velocity distributions of
substructures in dark halos. We also compare our analytic results with 
very recent high-resolution numerical simulations presented by
\citet{delucia04}.  

The plan of our paper is as follows. In \S \ref{sec:suv}, we study the
effects of global tides and dynamical frictions on the evolution of dark
halo substructures, and provide approximation formula to quantify them. 
Section \ref{sec:der} is devoted to present an analytic formalism
for the spatial and mass distributions of substructures. We
summarize the results in \S \ref{sec:result}, and we compare our analytic
predictions with numerical data in \S \ref{sec:comp}. Finally, in \S
\ref{sec:conc} we discuss the results and draw final conclusions.   

\section{GLOBAL TIDES AND DYNAMICAL FRICTIONS}
\label{sec:suv}

\subsection{Tidal Stripping}
\label{subsec:strip}
The most dominant force that drive the dynamical evolution of the dark
halo substructures ({\it subhalos}) are the global tides from host halos 
\citep{okamoto99}. The global tides from host halos strip the outer
parts of subhalos, resulting in the subhalo total disruption or at least 
significant amount of subhalo mass loss. 

For the realistic treatment of the mass-loss caused by the global tides, 
it is necessary to assign the density profiles to both the host 
halos and the subhalos. 
We consider a situation that a subhalo with with virial mass 
$m_{\rm vir}$ is moving in a circular orbit of radius $R$ from the 
center of a host halo with virial mass $M_{\rm vir}$. 
Then we assume that the initial density distributions of both the 
subhalos and the host halos are well described by the profiles 
obtained in N-body simulations \citep*[][hereafter NFW]{navarro97}:
 
\begin{equation}
 \rho(R)=\frac{\rho_s}{(R/R_{\rm s})(1+R/R_{\rm s})^2},
\label{nfw}
\end{equation}
where $\rho_{\rm s}$ is the characteristic density that can be computed 
from the nonlinear overdensity $\Delta_{\rm vir}$, and $R_{\rm s}$ is
the halo scale radius.  We use the top-hat radius as the virial radius
such that  
\begin{equation}
 R_{\rm vir} = \left(\frac{3M_{\rm vir}}
{4\pi\Delta_{\rm vir}\bar{\rho}}\right)^{1/3},
\end{equation}
where $\bar{\rho}$ is the mean density of the universe.
The ratio of the scale radius $R_{\rm s}$ to the halo virial
radius $R_{\rm vir}$ defines the halo concentration parameter $C$ and 
$M_{\rm vir}$ at redshift $z$ \citep{klypin99}: 
\begin{equation}
\label{concen}
 C=\frac{124}{1+z}\left(\frac{M_{\rm vir}}{1h^{-1}M_\odot}\right)^{-0.084},
\end{equation}
here we incorporate the redshift dependence found by e.g.,
\citet{bullock01}. 
The halo mass confined within the radius of $R$ can be computed from 
the following relation: 
\begin{equation}
\label{mr}
 M(R)=M_{\rm vir}\frac{f(X)}{f(C)},
\end{equation}
where
\begin{equation}
\label{fx}
 f(X) \equiv \ln(1+X)-\frac{X}{1+X},
\end{equation}
and $X \equiv R/R_{\rm s}$. Equations (\ref{nfw})-(\ref{fx}) are 
written in terms of the host halo properties but they also hold 
for the subhalo properties. From here on, the capital letters and 
the small letters are consistently used for the notation of host halos
and subhalos, respectively.  For example, $M$, $R$, and $C$ represent 
the mass, the radius, and the concentration parameter of host halos, while 
$m$, $r$, and $c$ are the same quantities for the subhalos.

The effect of global tide can be quantified by the tidal radius, 
$r_{\rm t}$, that is defined as the radius all the mass of a satellite 
beyond which gets lost by the global tides. For the realistic case that 
both the host halos and the subhalos are not point-like masses but have
extended profiles as is our case, the tidal radius $r_{\rm t}$ is given 
as 
\begin{equation}
\label{rt}
r_{\rm t} = {\rm min}(r_{\rm t0},r_{\rm re}),
\end{equation}
where $r_{\rm t0}$ and $r_{\rm re}$ are defined as a radius at which
gravity equals the tidal force and a radius determined by the
resonances, respectively. The following two equations allow us to
determine the values of $r_{\rm t0}$ and $r_{\rm re}$:
\begin{eqnarray}
\label{tidal1}
 \frac{f(x)}{f(X)}&=&\left(\frac{x}{X}\right)^3
\left(\frac{r_{\rm s}V_{\rm max}}{R_{\rm s}v_{\rm max}}\right)^2
\left[2-\frac{X^2}{(1+X)^2f(X)}\right], \\
 \frac{f(x)}{f(X)}&=&\left(\frac{x}{X}\right)^3\left(\frac{r_{\rm s}
V_{\rm max}}{R_{\rm s}v_{\rm max}}\right)^2,
\label{tidal2}
\end{eqnarray}
where $x \equiv r/r_{\rm s}$ and  $X \equiv R/R_{\rm s}$. 
Here $V_{\rm max}$ and $v_{\rm max}$ represent the maximum circular 
velocity of the host halo and the subhalo respectively. We approximate
that the maximum circular velocity occurs at twice the scale
radius \citep{klypin99}, $V_{\rm max}=V(2R_{\rm s})$ and $v_{\rm
max}=v(2r_{\rm s})$, where the rotation velocities are derived by
\begin{eqnarray}
\label{v1}
 V^2(R)&=&\frac{GM_{\rm vir}}{R}\frac{f(X)}{f(C)},\\
\label{v2}
 v^2(r)&=&\frac{Gm_{\rm vir}}{r}\frac{f(x)}{f(c)}.
\end{eqnarray}
Once the tidal radius is determined through the 
above equations (\ref{rt})-(\ref{v2}), the final subhalo mass, $m_{\rm f}$,
after the tidal stripping effect from the global tides is computed as 
\begin{equation}
\label{mfinal}
m_{\rm f} = m_{\rm vir}\frac{f(r_{\rm t}/r_{\rm s})}{f(c)},
\end{equation}
for $r_{\rm t}<r_{\rm vir}$ and $m_{\rm f}=m_{\rm vir}$ otherwise.

\subsection{Orbital Decay}
\label{subsec:decay}
Since the subhalos reside in very dense environments within the 
host halos, they undergo dynamical friction. Although the most dominant 
force that drives the subhalo mass-loss is the global tides from host
halos, dynamical friction also plays an important role in driving the
subhalo dynamical evolution. It causes the orbital decay of the
subhalos, making them more susceptible to strong tidal forces.

The key quantity in describing the orbital decay caused by dynamical
friction is the  friction time scale $t_{\rm df}$: 
\begin{equation}
 \frac{dR}{dt}=-\frac{R}{t_{\rm df}},
\end{equation}
which can be estimated by the Chandraskekhar's formula: 
\begin{eqnarray}
 t_{\rm df}(m_{\rm vir},R)&\!\!=\!\!&\frac{1}{2}\left[\frac{\partial\ln M(R)}
{\partial\ln R}+1\right]^{-1}\nonumber\\
&&\times\frac{V_{\rm circ}^3(R)}{4\pi G^2(\ln\Lambda)m_{\rm vir}
\rho(R)g(V_{\rm circ}(R)/\sqrt{2}\sigma_r)},
\label{fric_time}
\end{eqnarray}
where 
\begin{eqnarray}
\label{gxi}
 g(\xi) &\equiv& {\rm erf}(\xi)-\frac{2}{\sqrt{\pi}}\xi e^{-\xi^2}, \\
\label{lnL}
 \ln\Lambda&=&8,\\
\label{sigmar}
 \sigma_r^2&=&V_{\rm max}^2\frac{2X(1+X)^2}{f(2)}\int_{X}^\infty
\frac{f(x)}{x^3(1+x)^2}dx.
\end{eqnarray}
Here we assume that the initial circular velocities of the subhalos
follow the Maxwellian distributions \citep{binney87,klypin99}, and the 
Coulomb logarithm $\Lambda$ has a constant value of $8$ since this value 
is found to give the best result in simulations \citep{tormen98}. 

Since the dynamical friction timescale is approximately 
proportional to the radius $R$ \citep{klypin99}, the final decayed
radius $R_{\rm f}$ of the subhalo during the time interval $\Delta t$
can be estimated as  
\begin{equation}
\label{Rf}
R_{\rm f} = R_{\rm i}\left(1 - \frac{\Delta t}
{t_{\rm df}(m_{vir},R_{\rm i})}\right),
\end{equation}
where $R_{\rm i}$ is the initial orbital radius of the subhalo before
the orbital decay, and $m_{\rm vir}$ is the initial virial mass of the  
subhalo before the tidal stripping effect. From here on, we drop the 
subscript ''vir'' in denoting the initial virial mass, and use 
the notation $M$ and $m_{\rm i}$ for representing the virial
mass of host halos and the initial virial mass of subhalos,  
respectively.  

The model we adopt in this paper is rather simple, and differs from the
ones used in the recent comprehensive studies of these dynamical
effects. For instance, \citet{hayashi03} showed that the effects of tides
may be underestimated in the model described above, and that the impulse
approximation results in better agreement with the numerical simulations.
\citet{benson04} considered the improved dynamical friction model which
incorporates non-circular motions and more complicated Coulomb logarithm
$\Lambda$. Nevertheless we keep this simple model so as to be
analytically tractable. 

\section{DERIVATIONS OF SUBSTRUCTURE DISTRIBUTIONS}
\label{sec:der}

In \S \ref{sec:suv}, we have investigated the effects of global tides
and dynamical frictions on the evolution of the dark matter subhalos, and
showed that it is possible under some simplified assumptions to predict
the final mass and position of a subhalo from the initial mass and
position along with the given host halo mass.

The initial spatial and mass distribution of subhalos can be computed by
modifying the popular \citet[][hereafre PS]{press74} approach \citep{lee04}. 
Let $n(m_{\rm i},R_{\rm i};M,z,z_{\rm i})dR_{\rm i}dm_{\rm i}$ represent 
the number density of subhalos formed at redshift $z_{\rm i}$ with mass 
in the range of $[m_{\rm i},m_{\rm i}+dm_{\rm i}]$ located in a spherical 
shell of radius $R_{\rm i}$ with thickness of $dR_{\rm i}$ from the
center of a host halo with mass $M$ at redshift $z$. 
This number density can be written as a product of two distributions:
\begin{equation}
n(m_{\rm i},R_{\rm i};M,z,z_{\rm i}) = 
P(R_{\rm i};M,z)n(m_{\rm i}|R_{\rm i};M,z,z_{\rm i}),
\end{equation}
where $P(R_{\rm i};M,z)dR_{\rm i}$ is the probability that the subhalo has 
an orbital radius of $R_{\rm i}$ provided that it is included in a host
halo of mass $M$ at redshift $z$, and $n(m_{\rm i}|R_{\rm i};M,z,z_{\rm
i})dm_{\rm i}$ is the number density of subhalos formed at $z_{\rm i}$ with
initial mass in the range of $[m_{\rm i},m_{\rm i}+dm_{\rm i}]$ 
provided that it is located at an initial orbital radius of 
$R_{\rm i}$ from the center of a host halo of mass $M$ at redshift $z$.

For the probability $P(R_{\rm i};M,z)$, we make {\it apriori} assumption 
that it has a form of the NFW profile, expecting that the subhalos follow 
initially the distribution of dark matter particles:
\begin{equation}
 P(R_{\rm i}; M, z)dR_{\rm i}=
\frac{4\pi R_{\rm i}^2A}{(R_{\rm i}/R_{\rm s})(1+R_{\rm i}/R_{\rm s})^2}
dR_{\rm i},
\label{dist_spa}
\end{equation}
with
\begin{equation}
\label{A}
A\equiv \frac{1}{4\pi R_{\rm s}^3f(C)}.
\end{equation}
Here the amplitude $A$ was determined from the normalization constraint
of 
\begin{equation}
 \int_0^{R_{\rm vir}}P(R_{\rm i}; M, z)dR_{\rm i}=1.
\label{pr_norm}
\end{equation}
While we normalize the distribution by equation (\ref{pr_norm}),
we extend this distribution to $R_{\rm c}$, where the radius upper limit
$R_{\rm c}$ represents the effective range of the dynamical friction
force beyond which the  force is negligible. Since the host halo has an
extended density profile (see eq. [\ref{nfw}]), the effective range of
dynamical friction $R_{\rm c}$ does not necessarily coincide with the
virial radius of the host halo. Hence in order to derive the subhalo
distribution, we have to consider those subhalos which were initially
placed outside the virial radius of the host halo but eventually fell
into within the host halo virial radius due to the orbital decay caused
by the dynamical friction. We estimate the effective radius of the host
halo dynamical friction, to find $R_{\rm c} \approx 100R_{\rm s}$.
Hereafter we use the value of $R_{\rm c} = 10R_{\rm vir}$ since $R_{\rm
vir} \approx 10R_{\rm s}$.  As we will see in \S \ref{sec:result},
however, our results are insensitive to the specific choice of $R_{\rm c}$.

The conditional distribution, $n(m_{\rm i}|R_{\rm i};M,z,z_{\rm i})$, can
be obtained by incorporating the spatial correlation between the subhalo
and the host halo into the conditional PS mass function \citep*{yano96,lee04}:
\begin{equation}
 n(m_{\rm i}|R_{\rm i}; M, z, z_{\rm i})dm_{\rm i}=
\!\!\sqrt{\frac{2}{\pi}}\frac{M}{m_{\rm i}}
\left|\frac{\partial \sigma_{\rm s}}{\partial m_{\rm i}}
\right|\left|\frac{\partial\beta}{\partial\sigma_{\rm s}}
\right|e^{-\beta^2/2}dm_{\rm i},
\label{dist_mass}
\end{equation}
where
\begin{equation}
 \beta\equiv \frac{\delta_{\rm c}(z_{\rm i})}
{\sqrt{\sigma_{\rm s}^2-\sigma_{\rm c}^4/\sigma_{\rm h}^2}}
\left(1-\frac{\delta_{\rm c}(z)}{\delta_{\rm c}(z_{\rm i})}
\frac{\sigma_{\rm c}^2}{\sigma_{\rm h}^2}\right),
\end{equation}
\begin{equation}
 \sigma_{\rm s}^2\equiv\int_{-\infty}^{k(m_{\rm i})}\Delta^2(k)d\ln k,
\end{equation}
\begin{equation}
 \sigma_{\rm h}^2\equiv\int_{-\infty}^{k(M)}\Delta^2(k)d\ln k,
\end{equation}
\begin{equation}
 \sigma_{\rm c}^2\equiv\int_{-\infty}^{k(M)}\Delta^2(k)\frac{\sin kR_{\rm i}}
{kR_{\rm i}}d\ln k,
\end{equation}
where $\Delta(k)$ is the dimensionless power spectrum of the linear
density field, the wave number $k(M)$ is related to the mass as 
\begin{equation}
 k(M)=\left(\frac{6\pi^2\bar{\rho}}{M}\right)^{1/3},
\end{equation}
and $\delta_{\rm c}(z)$ is a threshold value of the dimensionless
density contrast $\delta$ at redshift $z$ given by $\delta_{\rm
c}(z)\approx 1.68/D(z)$, where $D(z)$ is the linear growth rate of the
density field. For an Einstein De-Sitter universe, 
it reduces to  $\delta_{\rm c}(z)\approx 1.68(1+z)$.  

Our use of a NFW profile form for the Lagrangian distribution, 
$P(R_{\rm i})$ can be justified as follows:  The original PS formalism 
assumes that bound objects form at the local density peaks. Thus, in 
the strict PS framework, a Gaussian peak profile \citep{bardeen86} 
has to be used for $P(R_{\rm i})$ to be consistent. 
However, several numerical simulations \citep*[e.g.,][]{katz93} have 
found a poor correlation between the location of bound objects and the local 
density peaks in Lagrangian space.  In other words, in spite of the fact that 
the PS theory happens to be quite successful in estimating the statistically 
averaged number density of bound objects, it has been shown to fail in 
predicting the Lagrangian density profiles of bound objects, $P(R_{\rm i})$.   
All one can say for sure is that no matter what the functional form 
of $P(R_{\rm i})$ is, when it is mapped to the Eulerian space, 
it must be close to the NFW profile.  
Without knowing a correct form of $P(R_{\rm i})$, a possible way to make 
a best approximation is to use the form of NFW profile itself in Lagrangian 
space\footnote{Since we are considering a region which has yet to
virialize, it may have a different profile from the NFW profile; for
instance, \citet{sheth01} considered a time-evolution of such profiles by
modifying the scale radius.}. This guarantees that the corresponding
Eulerian profile is NFW, and allows us to use the PS approach in
Lagrangian space to the substructure mass function. 

The other quantity to be considered is the subhalo formation
epoch\footnote{The formation epoch is defined as the epoch when a
progenitor with a mass of $fM$ is formed. We adopt $f = 1/2$
throughout this paper, which is a standard value used by \citet{lacey93}
and in many studies which is a standard choice in this context. 
We note that our result is rather insensitive to the specific choice of
$f$.}
distribution, $dp/dz$, since the orbital decay of subhalos depend on the
time interval $\Delta t \equiv t - t_{\rm i}$. \citet{lacey93} derived an
analytic expression for $dp/dz$, and \citet{kitayama96} provided a 
fitting formula to $dp/dz$, which we adopt here 
\citep[see Appendix C in][]{kitayama96}. 

\begin{figure*}
  \begin{center}
  \includegraphics[width=0.8\hsize]{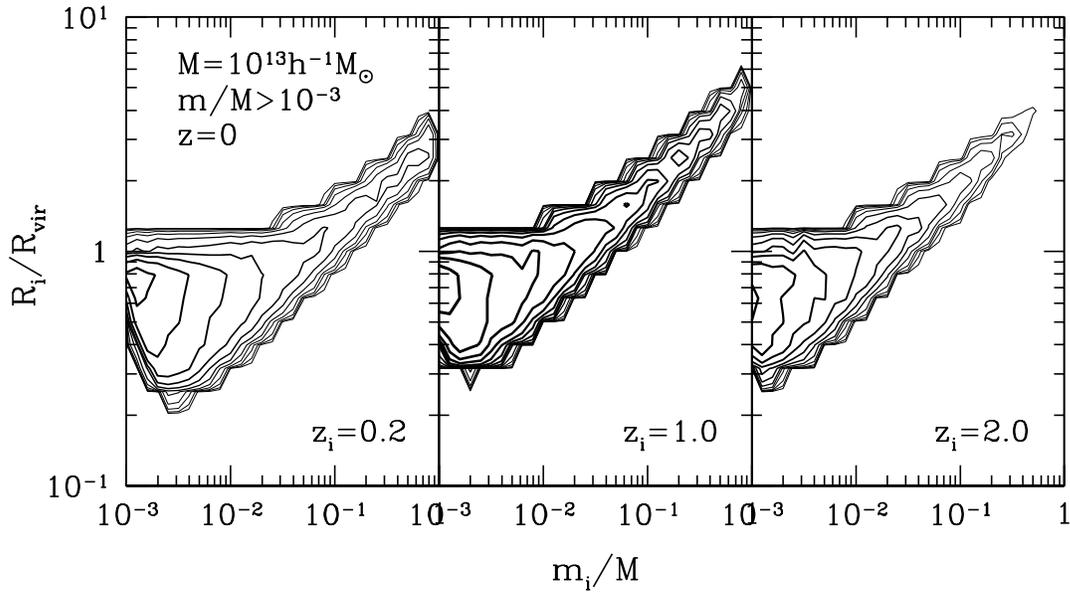}
   \end{center}
    \caption{The region of the integral (eq. [\ref{join_dis}]) in the
 $m_{\rm i}$-$R_{\rm i}$ plane, for fixed values of $z_{\rm i}$; 
$z_{\rm i}=0.2$ ({\it left}), $1.0$ ({\it middle}), and $2.0$ ({\it
 right}). We plot contours of the integrand (see eq. [\ref{join_dis}])
 in the logarithmic space; 
 $(dp/dz_{\rm i})n(m_{\rm i},R_{\rm i};M,z,z_{\rm i})m_{\rm i}R_{\rm i}$. 
 Darker contours mean larger values, thus darker regions contribute to
 the integral more.} 
\label{fig:mc}
\end{figure*}

The analytic steps to reach the final subhalo spatial and mass
distributions with taking account of tidal mass-loss and orbital decay 
caused by dynamical friction are summarized as follows:
\begin{enumerate}
\item We determine the final position $R_{\rm f}$ of a subhalo after the
      dynamical friction effect through equations
      (\ref{fric_time})-(\ref{Rf}). 
\item We determine the tidal radius $r_{\rm t}$ of the subhalo at the final 
position $R_{\rm f}$ through equations (\ref{rt})-(\ref{v2}).
\item We abandon those subhalos whose tidal radius $r_{\rm t}$ is larger  
than its final orbital radius $R_{\rm f}$, assuming that they will 
      get disrupted completely by the tidal stripping effect. 
\item For the survived subhalos with $r_{\rm t} < R_{\rm f}$, we
      determine the final mass $m_{\rm f}$ through equation (\ref{mfinal}).
\item Finally, we count the cumulative number of those subhalos as
      function of mass $m$ and radius $R$.
\end{enumerate}
The above procedures to evaluate the cumulative spatial and mass 
distributions of the subhalos can be summarized by the following
analytical expression:
\begin{eqnarray}
\label{join_dis}
N(>m,>R;M,z) &=& \int \frac{dp}{dz_{\rm i}} dz_{\rm i} \int_{S}dm_{\rm i}dR_{\rm i} \nonumber\\
&&\times n(m_{\rm i},R_{\rm i};M,z,z_{\rm i}).
\end{eqnarray}
where $S$ represents the following condition: 
\begin{eqnarray}
m_{\rm f}(m_{\rm i},R_{\rm i},z_{\rm i})& >& m,\label{condition1}\\
R_{\rm f}(m_{\rm i},R_{\rm i},z_{\rm i})& >& R,\label{condition2}\\
R_{\rm vir}& >& R_{\rm f}(m_{\rm i},R_{\rm i},z_{\rm i}), \label{condition3}\\
R_{\rm f}(m_{\rm i},R_{\rm i},z_{\rm i})& >& r_{\rm t}
(m_{\rm i},R_{\rm i},z_{\rm i}),\label{condition4}
\end{eqnarray}
In other words, the integration over $dm_{\rm i}dR_{\rm i}$ is performed 
only those regions in the $m_{\rm i}$-$R_{\rm i}$ plane that satisfy the 
above conditions (eqs. [\ref{condition1}]-[\ref{condition4}]). To
perform an integral with such a complicated boundary condition, it
is useful to use a Monte-Carlo integral method. 

\section{RESULT}
\label{sec:result}

As a specific example, we show results of our analytic model in a Lambda
dominated CDM model with the choice of the mass density parameter
$\Omega_{\rm m}=0.3$, the vacuum energy density parameter
$\Omega_{\Lambda}=0.7$, the spectral shape parameter $\Gamma=0.168$, the
dimensionless Hubble constant of $h=0.7$, and the rms fluctuation
normalization $\sigma_8=0.9$. We assume the power spectrum of cold dark
matter model with primordial spectral index $n_{\rm i}=1$, and adopt a
fitting formula of \citet{bardeen86}. 

In Figure \ref{fig:mc}, we show the integral region, which is determined
by equations (\ref{condition1})-(\ref{condition4}), in the $m_{\rm
i}$-$R_{\rm i}$ plane for fixed values of $z_{\rm i}$. In reality, we
plot contours of the integrand (see eq. [\ref{join_dis}]) in the
logarithmic space. As seen, the region of the integral has a
simple topology, and is easily understood; the more massive
substructures tend to sink faster, thus they should be in the outer
regions of the host halo; when $z_{\rm i}$ is larger the typical value
of $R_{\rm i}$ also becomes larger because substructures experience
dynamical friction more. However, we also find that the contribution 
of massive substructures to the integral is rather minor. This is simply
because of the small number of massive substructures. Too small $R_{\rm
i}$ is not allowed because final sizes of substructures should be
smaller than the final distances from the center (eq.
[\ref{condition4}]).  It is also clear from this Figure that  our
results are insensitive to the value of $R_{\rm c}$ as far as we adopt
sufficiently large $R_{\rm c}$, $R_{\rm c}> 5R_{\rm vir}$, since
substructures which initially lie very far from the host halo are not
counted after all.   

\subsection{Mass Distribution}

The cumulative mass distribution $N(>m;M,z)$ of subhalos in a given 
host halo of mass $M$ at redshift $z$ is now straightforwardly obtained 
by setting $R = 0$ in equation (\ref{join_dis}):
\begin{equation}
\label{mdis}
N(>m;M,z) \equiv N(>m,>0;M,z).
\end{equation}

\begin{figure}
  \begin{center}
   \includegraphics[width=0.8\hsize]{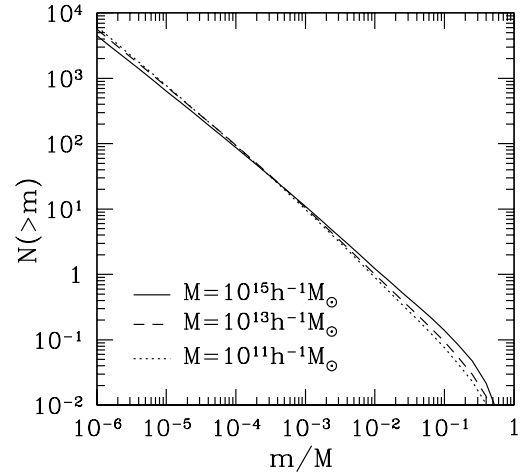}
  \end{center}
    \caption{Cumulative mass function of substructures in units of
 rescaled substructure mass. The redshift is $z=0$.
 For the mass of the host halo, we consider 
 $M=10^{15}h^{-1}M_\odot$ ({\it solid}), 
 $10^{13}h^{-1}M_\odot$ ({\it dashed}), 
 and $10^{11}h^{-1}M_\odot$ ({\it dotted}). }
\label{fig:cum_mass}
\end{figure}

   \begin{figure*}
  \begin{center}
  \includegraphics[width=0.8\hsize]{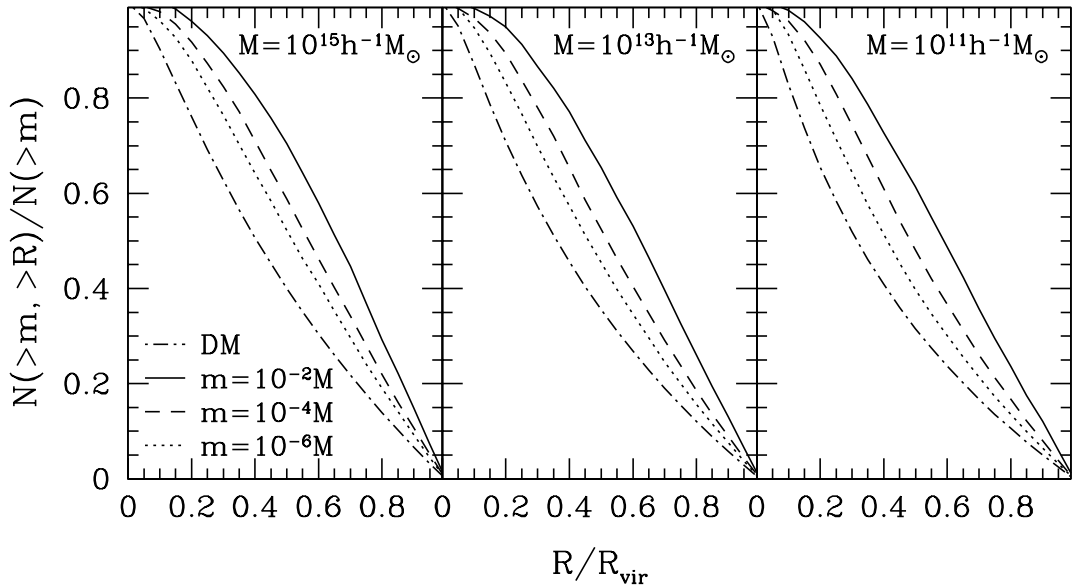}
    \end{center}
   \caption{Normalized cumulative spatial distribution of substructures
 in units of  rescaled radius. Mass of the host halo is set to 
 $M=10^{15}h^{-1}M_\odot$ ({\it left}), 
 $10^{13}h^{-1}M_\odot$ ({\it middle}), 
 and $10^{11}h^{-1}M_\odot$ ({\it right}).  We only consider
 substructures with mass larger than $10^{-2}M$ ({\it solid}),
 $10^{-4}M$ ({\it dashed}), and $10^{-6}M$ ({\it
 dotted}). The spatial distribution of smooth dark matter component,
 which can be calculated from equation (\ref{nfw}), is also shown by
 dash-dotted lines. }
\label{fig:cum_rad}
   \end{figure*}

Figure \ref{fig:cum_mass} plots $N(>m;M,z)$ at redshift $z=0$ as a
function of rescaled subhalo mass for the three different cases of host
halo mass $M$. Note that all the three curves in Figure
\ref{fig:cum_mass} are very similar to one 
another, indicating that the subhalo mass distribution is almost
independent of the host halo mass. In addition, these curves are close to
a power law with the slope of $\sim -0.9$ in low mass range of 
$m/M \sim 10^{-5}$ and $\sim -1.0$ for high mass range of 
$m/M \sim 10^{-2}$. These findings from our analytic model are all
consistent with recent high-resolution numerical simulation results 
\citep[e.g.,][]{ghigna00,delucia04}. It is worth noticing, however, that
there is some tendency that the larger the host halos are the slightly 
more massive subhalos they are assigned to, which is likely because
larger host halos form relatively late so that they have relatively
little time for their subhalos to undergo orbital decay and lose mass.

\subsection{Spatial Distribution}

Equation (\ref{join_dis}) can be interpreted as the cumulative spatial
distribution of subhalos for a given mass range. 
Figure \ref{fig:cum_rad} plots the normalized cumulative spatial
distribution of subhalos as a function of rescaled radius, and the 
spatial distributions of the smooth component dark matter for comparison
as well. The normalized spatial distribution is defined as
$N(>m,>R;M,z)/N(>m;M,z)$.  One can see that the spatial 
distributions of the dark subhalos are {\it anti-biased} relative 
to that of the dark matter particle components. This phenomenon has been
observed in several numerical simulations
\citep*[e.g.,][]{ghigna00,delucia04}. We also note that more massive 
subhalos have stronger tendency of anti-bias, which is likely because  
this anti-bias is caused by tidal stripping and dynamical friction, so
that massive subhalos are more affected by those dynamical processes.
 
\subsection{Velocity Distribution}

\begin{figure}
  \begin{center}
  \includegraphics[width=0.8\hsize]{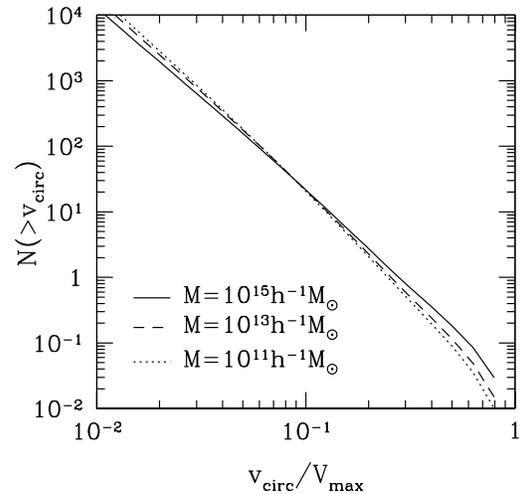}
  \end{center}
    \caption{Cumulative circular velocity function of substructures in
 units of rescaled substructure velocity. For the mass of the host halo,
 we consider  $M=10^{15}h^{-1}M_\odot$ ({\it solid}), 
 $10^{13}h^{-1}M_\odot$ ({\it dashed}), 
 and $10^{11}h^{-1}M_\odot$ ({\it dotted}). }
\label{fig:cum_vel}
\end{figure}

The circular velocity distribution of subhalos is sometimes more
useful in practice  because it is more readily to be compared with 
observations \citep{cole89,shimasaku93,gonzalez00,sheth03c,desai04}. 
Our analytic model also enables us to compute the circular velocity 
distribution. Without the tidal stripping effect, 
the subhalo would retain the original NFW density profile, and thus 
the subhalo circular velocity would be given by $v_{\rm max}=v(2r_{\rm s})$. 
However, because of the tidal stripping effect, the subhalos eventually 
end up with having truncated density profile at truncation radius 
$\sim 2r_{\rm s}$  \citep{hayashi03}. Thus, in
our realistic model, the subhalo circular velocity is given as  
\begin{eqnarray}
\label{circular}
  v_{\rm circ} = \left\{
      \begin{array}{ll}
        v(2r_{\rm s}; m_{\rm i}, z_{\rm i}) &
        \mbox{($r_{\rm t}>2r_{\rm s}$)}, \\ 
        v(r_{\rm t}; m_{\rm i}, z_{\rm i}) &
        \mbox{($r_{\rm t}<2r_{\rm s}$)}. 
      \end{array}
   \right. 
\end{eqnarray}
Thus, the subhalo circular velocity depends not only on the subhalo 
mass but also on the subhalo position and formation epoch. 
Using this relation (eq. [\ref{circular}]) along with the subhalo 
spatial and mass distribution (eq. [\ref{join_dis}]), we
can also evaluate the subhalo velocity distribution. 

Figure \ref{fig:cum_vel} plots the cumulative circular velocity
distribution of subhalos versus the subhalo velocity rescaled by 
$V_{\rm max}$. Note that the subhalo velocity distribution also shows 
very weak dependence on the host halo mass, power-law shaped with the 
slope of $\sim -3$ around $v_{\rm circ}/V_{\rm max}\sim10^{-1}$, which 
is all consistent with numerical detections \citep{ghigna00}. 
And it also has the same tendency as the spatial distribution that 
the larger a host halo is the more subhalos with high velocity it has. 

\section{COMPARISON WITH NUMERICAL SIMULATIONS}
\label{sec:comp}

\begin{figure*}
  \begin{center}
   \includegraphics[width=0.8\hsize]{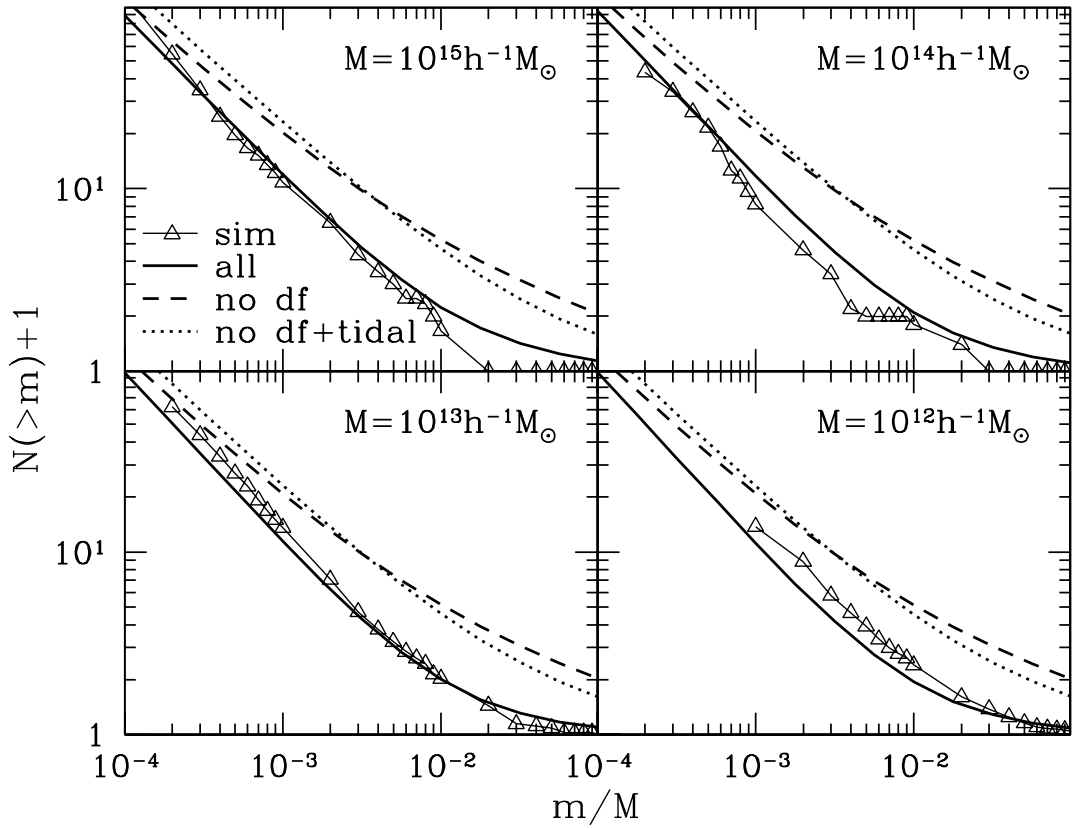}
  \end{center}
    \caption{Comparison of the analytic mass distributions ({\it thick lines})
 with numerical ones ({\it thin lines and triangles}) computed by
 \citet{delucia04}. We consider the host halo mass of $M=10^{15}h^{-1}M_\odot$ ({\it
 upper left}), $10^{14}h^{-1}M_\odot$ ({\it upper right}),
 $10^{13}h^{-1}M_\odot$ ({\it lower left}), and $10^{12}h^{-1}M_\odot$
 ({\it lower right}). As for the analytic mass distributions, we
 consider not only the fiducial mass distributions in which both
 dynamical friction and tidal stripping are included ({\it solid}), but
 also the mass distributions without dynamical friction ({\it dashed})
 and without dynamical friction and tidal stripping ({\it dotted}).
 Here we use $N(m)+1$ rather than $N(>m)$ because
 in Figure 1e of \citet{delucia04} the host halo itself was included in the
 cumulative mass function.} 
\label{fig:sim_mass}
\end{figure*}

\begin{figure*}
  \begin{center}
   \includegraphics[width=0.8\hsize]{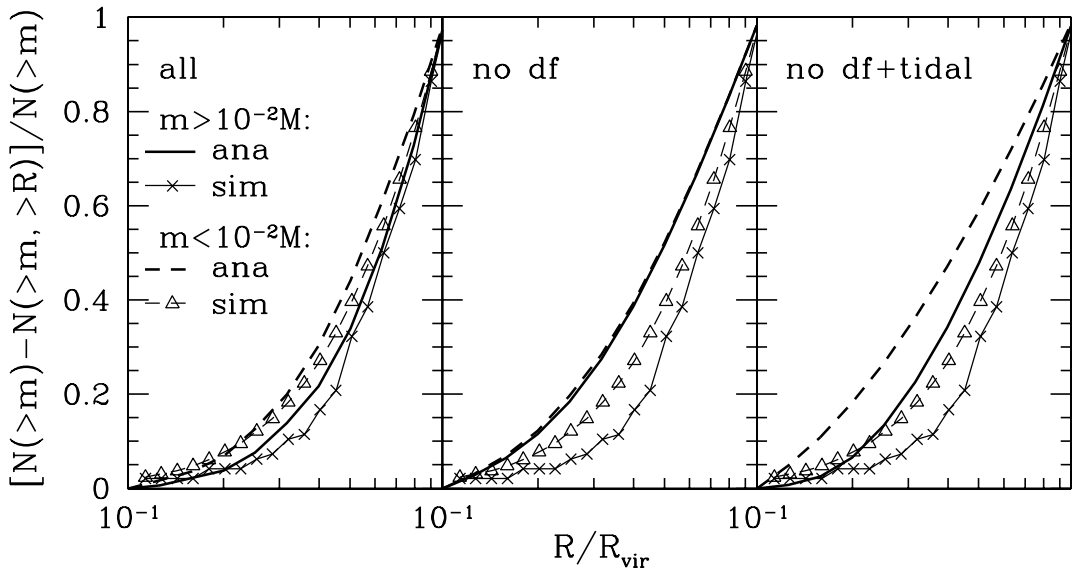}
  \end{center}
    \caption{Comparison of the analytic spatial distributions ({\it
 thick lines}) with numerical ones ({\it thin lines and symbols})
 computed by \citet{delucia04}. Comparison is done for both massive
 substructures ($m>10^{-2}M$, denoted by solid lines and/or crosses)  
 and less massive substructures ($10^{-2}M>m>4\times10^{-4}M$, denoted
 by dashed lines and/or triangles). From left to right panels, we compare the  
 fiducial analytic model, the model without dynamical friction, and the
 model without dynamical friction and tidal stripping.  } 
\label{fig:sim_rad}
\end{figure*}

In this section, we compare our analytic predictions with numerical data
from high-resolution simulations performed by \citet{delucia04}. 
In addition to our own analytic model, we also compare the following 
two models with numerical simulations to demonstrate the importance of
dynamical processes that we have included: (i)
the model with tidal stripping but without dynamical friction. (ii) the
model without tidal stripping  and dynamical friction. The models (i)
and (ii) basically correspond to those  considered in \citet{lee04} and
\citet{fujita02}, respectively. 

First in Figure \ref{fig:sim_mass} we plot our analytic predictions of the
subhalo mass distribution with the numerical simulation results
(see Fig. 1e of \citealt{delucia04}). We find that our model agrees 
quite well with the numerical simulations, while the models without dynamical
friction or tidal stripping effect tend to overpredict the number of massive
substructures. Those models are inconsistent with the numerical
simulations even if we allow them to change their normalization. The
success of our model indicates that the effects of tidal
stripping and dynamical friction are essential to understand the
evolution of subhalos and understand their distributions. 
For $M=10^{15}h^{-1}M_\odot$ and
$10^{14}h^{-1}M_\odot$, it seems that there is a small difference at
$m/M>0.02$. However, this difference is not significant
given the small numbers of host halos ($5$ and $6$, respectively) 
used for studying the mass distribution in the numerical simulations. 

In Figure \ref{fig:sim_rad} we compare our analytic predictions 
on the subhalo spatial distribution with numerical data 
(see Figure 7 of \citealt{delucia04}), and show that the two results
agree with each other quite well. It is clear from this Figure that our
model successfully reproduces the feature found in the numerical simulations
that more massive substructures are preferentially located in the outer
part of their host halo. This is not the case with the model without
dynamical friction. The model without tidal stripping and dynamical
friction also shows this feature, because we have assumed also in this
model that those subhalos whose radius $r_{\rm vir}$ is larger  
than its final orbital radius $R_{\rm f}$ will get disrupted completely.
However, the distributions themselves are largely inconsistent with the
numerical distributions, and rather similar to the distribution of dark
matter particles as for smaller substructures. It implies that the
dynamical frictions is the main reason for the anti-bias of the subhalo
spatial distribution relative to that of the dark matter particles.

\section{DISCUSSIONS AND CONCLUSIONS}
\label{sec:conc}

We have constructed an analytic model for mass and spatial distributions
of dark subhalos by taking account of two dominant dynamical processes 
that drive dominantly subhalo evolution: 
one is tidal stripping and consequent mass-loss caused by 
the host halo global tides which remove the outer, and the
other is the orbital decay caused by dynamical friction 
which drives massive subhalos to the inner part of the host halo. 
We also incorporate the formation epoch variation of the host halo, 
and the orbital decay of satellite halos outside the host halo virial
radius.  

We have found that our model predicts nearly power-law mass distribution
with weak dependence of host halo mass. The slope of the cumulative mass
function turned out to be about $-0.9$ for low mass substructures
$m/M \sim 10^{-5}$, and about $-1.0$ for larger mass
substructures $m/M \sim 10^{-2}$. Next we have predicted
the spatial distributions for given mass ranges of substructures. We have
found that the spatial distributions are basically anti-biased relative
to the smooth component, i.e., the substructures are preferentially
located in the outer part of the host halo. The extent of the anti-bias
is larger for more massive subhalos. Using our model, we have also
predicted the velocity distributions of subhalos. These findings are 
all consistent with what has been detected by recent high-resolution 
numerical simulation.  So, we have provided physical explanations to 
those characteristic findings of numerical simulations on the subhalo 
distributions.
 
There is some discussion about the accuracy of the \citet{bullock01}
choice of the evolution of the concentration parameter. 
For instance, \citet{zhao03} found that all high-redshift massive halos
have a similar median concentration, $C\sim 3.5$. To see how this
affects our results, we repeat the computation by assuming the
concentration parameter of equation (\ref{concen}) if $C\geq 3.5$, and
replace it with $C=3.5$ if equation (\ref{concen}) yields $C<3.5$.
We find that the substructure mass functions are not changed with this
replacement; at $z<1$ the change is unnoticeable, and even at $z=2$ it
changes the substructure mass function by $<10$\%.

To test the validity of our model, we have compared our analytic model
with the results of the high-resolution numerical simulations by
\citet{delucia04}. We have showed that mass and spatial distributions of
substructures in our model excellently agrees with those in the numerical
simulations.  For comparison, we have considered the models without
tidal stripping or dynamical friction and found that they cannot
reproduce the distributions in the numerical simulations.  This indicates
that the effects of both tidal stripping and dynamical friction are
very essential for constructing a realistic model of subhalo
distributions. 

Yet, it should be noted that we have adopted several simplified assumptions. 
For instance, we have ignored the effect of encounters between
substructures, although it may lead to mass loss comparable to that
caused by global tidal stripping \citep{tormen98}. We have also
assumed spherical halos and circular orbits of subhalos, which are 
inaccurate because halos in the CDM model are shown to be rather triaxial
\citep{jing02} and the subhalo orbits are shown to be quite 
eccentric \citep{tormen98,hayashi03}. In spite of these drawbacks of
our analytic model,  we believe that our model for the subhalo
distributions is the most realistic one that has ever been suggested,
and that it can be applied to wide fields such as gravitational lensing
and halo model of large-scale structure. Our analytic model is quite
useful also in calculating the mass and spatial distributions of
substructures for different initial conditions such as running primordial
power spectrum suggested by the combined analysis of cosmic microwave
background and large-scale structure observations \citep{spergel03}. 

\section*{Acknowledgments}
We are very grateful to Gabriella De Lucia for providing numerical 
simulation data for us, and for many useful suggestions. 
We also thank an anonymous referee for useful comments and suggestions. 
M. O. and J. L. acknowledges gratefully the research grant of the JSPS 
(Japan Society of Promotion of Science) fellowship. 


\label{lastpage}

\end{document}